\documentclass[12pt,onecolumn]{IEEEtran}
\usepackage[colorlinks,bookmarksopen,bookmarksnumbered,citecolor=red,urlcolor=red,]{hyperref}
\usepackage{indentfirst}
\usepackage{algorithm,algorithmic,amsbsy,amsmath,amssymb,epsfig,bbm,mathrsfs, bbm,tabu} 
\usepackage{multirow}
\usepackage{mathrsfs}
\usepackage{epstopdf}

\usepackage{verbatim}
\usepackage{amsfonts}
\usepackage{amsthm}
\begin{document}

\title{Smart Data Pricing Models for Internet-of-Things (IoT): A Bundling Strategy Approach}
\author{Dusit Niyato, Dinh Thai Hoang, Nguyen Cong Luong, Ping Wang, Dong In Kim, and Zhu Han
\thanks{$^*$ Dong In Kim is the corresponding author (e-mail: dikim@skku.ac.kr).}}

\maketitle

\begin{abstract}
Internet of things (IoT) has emerged as a new paradigm for the future Internet. In IoT, enormous devices are connected to the Internet and thereby being a huge data source for numerous applications. In this article, we focus on addressing data management in IoT through using a smart data pricing (SDP) approach. With SDP, data can be managed flexibly and efficiently through intelligent and adaptive incentive mechanisms. Moreover, it is a major source of revenue for providers and partners. We propose a new pricing scheme for IoT service providers to determine the sensing data buying price and IoT service subscription fee offered to sensor owners and service users, respectively. Additionally, we adopt the bundling strategy that allows multiple providers to form a coalition and bid their services as a bundle, attracting more users and achieving higher revenue. Finally, we outline some important open research issues for SDP and IoT.
\end{abstract}


\section{Introduction}

Internet of things (IoT) is a novel concept that allows a number of devices to be connected through the Internet. Such devices can be sensors/actuators which are able to operate and transmit data without or with minimal human intervention. IoT has brought a great influence to many areas, and there have been many IoT applications implemented such as healthcare, transportation, logistics, and manufacturing~\cite{Gubbi2013Internet}. However, the development of IoT is facing many challenges especially for data management~\cite{Bucherer2011Business}. Due to the special characteristics of IoT systems and services, e.g., heterogeneous large-scale architecture, diverse and enormous data, traditional data management approaches may become intractable such that new solutions are required. 

Recently, the concept of \emph{smart data pricing (SDP)}~\cite{Sen_2014_Smart} has been introduced as an alterative to address network resource management issues. Its major benefit is the ability to provide effective solutions from both system and economic aspects. In particular, with SDP, prices are used not only to gain profits for providers, but also to provide tools to improve system and data management. In this article, we propose to apply SDP to IoT so that prices are used to incentivize sensor owners to contribute their data to IoT services, improving the service quality and generating higher revenue from selling IoT services to users.

We first present an overview of IoT including its intrinsic features, system architecture, benefits, data management, and business model. We then introduce the SDP and review some related work. Moreover, we demonstrate the applications of economic models in IoT by introducing a pricing scheme to optimize the sensing data buying price and IoT service subscription fee for sensor owners and service users, respectively. We adopt the bundling strategy for multiple IoT providers to form a coalition and offer their services as a bundle. With bundling, the profit of the providers can be improved by encouraging many users to subscribe more services. Finally, we highlight a few important future research directions.


\section{An Overview of Internet of Things (IoT)}
\label{sec:Overview}

\subsection{Definitions and Features of IoT}
\label{subsec:}

Though the technology development and applications of IoT are tremendously growing, its definition can be diverse and fuzzy. \cite{IEEE_IoT_definition} provides one of the formal, concrete, and standardized definitions of IoT: ``\emph{Internet of Things envisions a self-configuring, adaptive, complex network that interconnects `things' to the Internet through the use of standard communication protocols. The interconnected things have physical or virtual representation in the digital world, sensing/actuation capability, a programmability feature and are uniquely identifiable.}''

In the IoT context, things can be objects that have Internet capability. The objects with unique identification can offer services in terms of data capture, communication, and actuation. Thus, derived from the definition, the fundamental features of IoT are as follows~\cite{Stankovic2014}:
\begin{itemize}
\item \emph{Connected to the Internet:} Things must be collected to the Internet using wired or wireless connections.
\item \emph{Uniquely:} Things are uniquely identifiable via IP addresses.
\item \emph{Sensing/Actuation capability:} Things are able to perform sensing/actuation tasks.
\item \emph{Embedded intelligence:} Things are embedded with intelligent functions, e.g., self-configurability.
\item \emph{Interoperable communication capability:} The IoT system has a communication capability based on standard technologies.
\end{itemize}

\subsection{Architecture of IoT}
\label{subsec:}

\begin{figure}
\begin{center}
$\begin{array}{c} \epsfxsize=4.3 in \epsffile{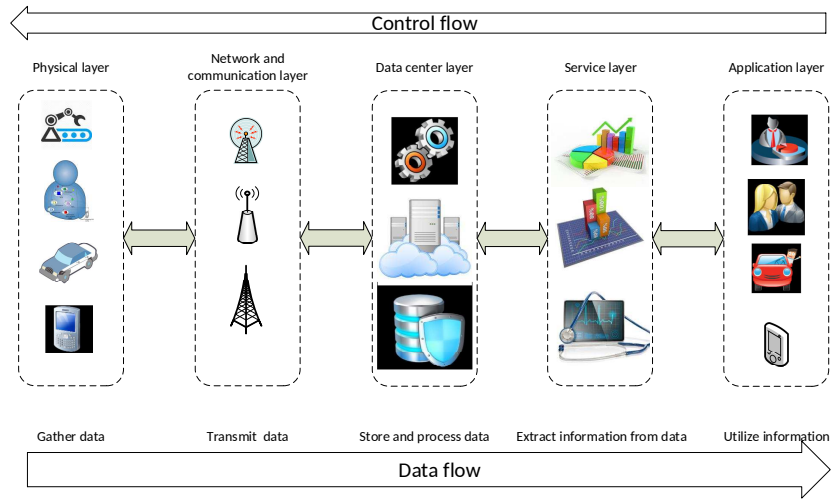} \\ [-0.2cm]
\end{array}$
\caption{A general architecture of IoT system.}
\label{IoT_architecture}
\end{center}
\end{figure}

A general architecture for IoT systems can be illustrated as in Figure~\ref{IoT_architecture}. The architecture consists of four layers.
\begin{itemize}
\item \emph{Physical layer} is composed of smart devices, e.g., sensors and actuators, that interact and/or gather data from the physical entities, e.g., environment and human.
\item \emph{Network and communication layer} provides data communications and networking infrastructure to transfer data collected from devices at the physical layer to the data center.
\item \emph{Data center layer} provides infrastructures to support data storage and processing to meet requirements of IoT applications.
\item \emph{Service layer} is a set of software that provides services to IoT users.
\end{itemize}
As shown in Figure~\ref{IoT_architecture}, the interactions among layers in the IoT architecture is through data. Data is first gathered at the physical layer by sensors and then transmitted to the data center through the network and communication layer. At the data center layer, the data is stored and processed to extract useful information for users in the service layer. Based on the received information, the applications or users can make appropriate decisions, e.g., sending commands to control sensors/actuators at the physical layer. 

\subsection{Benefits of IoT}
\label{subsec:}

With an idea to connect everything to the Internet, IoT has brought a variety of benefits. The major benefits are to improve system efficiency and user satisfaction, enhance flexibility, safety, and security, and finally open new business opportunity and revenue stream.

Because of many benefits, there are numerous applications of IoT in various sectors such as education, economics, transportations, and healthcare~\cite{Gubbi2013Internet}. We discuss one example in the following. Tomtom (www.tomtom.com), a well known GPS manufacturer, introduces an IoT service for a congestion index. The aim is to capture anonymous travel time information particularly in urban areas. Data obtained from smart devices, e.g., vehicle journey recorders, GPS, and traffic cameras, will be sent to the Tomtom servers through communication channels such as 3G/WiFi. The data is used to extract meaningful traffic information. Based on the information, the Tomtom service provides travel time information and real-life driving patterns presented to the general public, industry, and policy makers showing global congestion level. Thus, using the Tomtom service, drivers can have more efficient and safe journeys. Government agencies and authorities can make appropriate policies, rules, and regulations in controlling road traffic, reducing accidents, and constructing road infrastructure. Finally, businesses can have useful information for their operations, e.g., to open retail stores, gas stations, and repair shops at the best locations.

\subsection{Data Management in IoT}
\label{subsec:}

As shown in the IoT architecture (Figure~\ref{IoT_architecture}), data is the key component, and hence data management is a main concern in IoT. Data management involves the following aspects.
\begin{itemize}
\item \emph{Data collection:} Sensing is the first step in the data flow to obtain IoT data. In the sensing process, to collect a large amount of data with high quality, a number of sensors have to be deployed which will result in high costs. Sensor deployment, data gathering and preprocessing have to be optimized. Alternatively, participatory sensing can be adopted.
\item \emph{Data communications:} Together with a large number of devices, data communication and networking become important issues that need further analysis and optimization to meet specific requirements of IoT. Machine-type-communications (MTC) has been introduced as the solutions for cellular networks. Alternatively, Bluetooth low energy (BLE), WiFi, and 6LoWPAN can be used for local and personal area networks.
\item \emph{Data storage and processing:} The amount of IoT data is enormous which needs to be stored and processed efficiently and securely. Cloud computing becomes an effective solution. 
\item \emph{Information trading:} After data is processed, the useful information will be extracted to provide (sell) services to IoT users. This is an important step in business models in terms of profitability and sustainability. Market structures, incentive and pricing mechanisms have to be newly defined because of specific characteristics of IoT systems, businesses, and users. 
\end{itemize}

\subsection{Business Models and IoT}
\label{subsec:Business_IoT}

\begin{figure}
\begin{center}
$\begin{array}{c} \epsfxsize=5.0 in \epsffile{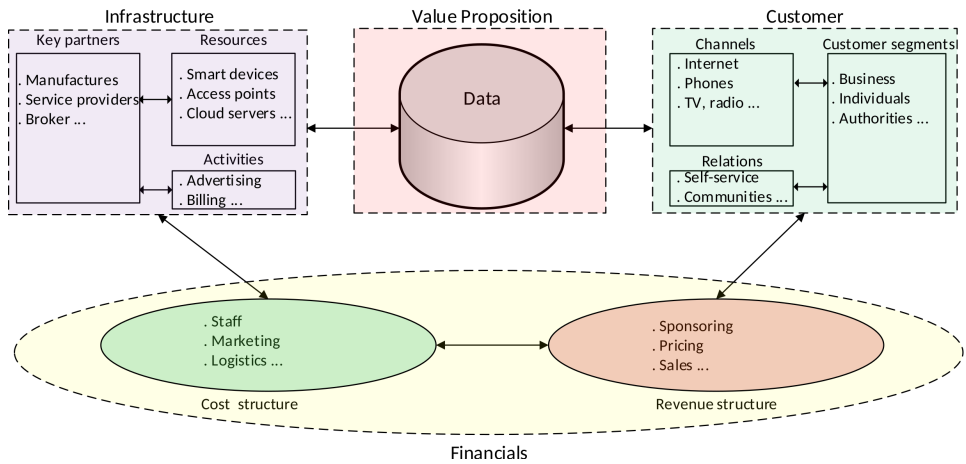} \\ [-0.2cm]
\end{array}$
\caption{A general business model of IoT systems.}
\label{Business_Model_Framework}
\end{center}
\end{figure}

The core economic benefit of IoT is to generate revenues for businesses, and thus an IoT business model is important. Generally, a business model \emph{describes the rationale of how a company creates, delivers, and captures value}~\cite{Osterwalder2010Business}. There are four major components in the business model as shown in Figure~\ref{Business_Model_Framework}. The \emph{value proposition} of data is the core component in the business model. The value proposition specifies what to be actually delivered and the price that will be charged to the customers. The main purpose of the value proposition is for the company to demonstrate to the customers that they will gain more benefits than what they pay. Accordingly, it is important in the business model to specify pricing mechanisms. To determine the price, the company needs to analyze and know the total cost incurred to operate the business and to offer IoT services. The major cost is from investing in and operating \emph{Infrastructure} components. Moreover, the company has to quantify the willingness-to-pay value of the customers. The economic and marketing techniques can be employed to understand the utility structure based on the types and segments of the customers. With the information about cost and willingness-to-pay, the company can derive an optimal price that maximizes the provider profit and user satisfaction.

In the next section, we introduce and review pricing models developed for IoT data management.

\section{Smart Data Pricing (SDP) Approaches in IoT}
\label{sec:SmartData}

SDP~\cite{Sen_2014_Smart} is a new concept to enhance network performance and to support data management through using pricing incentives. For example, during a congestion period, dynamic pricing is used to defer some non-urgent users from accessing networks, improving QoS performance.

There are two important attributes that make SDP a suitable solution for the data management in IoT. 
\begin{itemize}
\item \emph{Flexibility:} Instead of using static pricing, e.g., the usage-based method (byte-counting) that requires users to pay the same price per unit of data accessed, SDP provides the flexible and dynamic pricing mechanisms based on the demand and requirements of the users. 
\item \emph{Incentive:} Devices may belong to different owners who may have no interest in the joint data management. Thus, the use of SDP can attract such independent device owners to participate in IoT, improving the service quality. 
\end{itemize}

In this section, we review related applications of SDP in IoT data management. The summary of related applications is given in Table~\ref{table_sec3_sum}.

\begin{table*}[h]
\caption{Summary of pricing models and key descriptions}
\label{table_sec3_sum}
\scriptsize 
\begin{centering}
\begin{tabular}{|>{\centering\arraybackslash}m{2cm}|>{\centering\arraybackslash}m{2cm}|>{\centering\arraybackslash}m{5.5cm}|>{\centering\arraybackslash}m{4cm}|>{\centering\arraybackslash}m{2cm}|}
\hline
\multirow{2}{*} {\textbf{Pricing model}}&\multirow{2}{*} {\textbf{Market structure}} &\multirow{2}{*} {\textbf{Key descriptions}} &\multirow{2}{*} {\textbf{Suitable scenarios}}  &\multirow{2}{*} {\textbf{Solution}} \tabularnewline 
& & & & \tabularnewline 
\hline
\hline
\vspace{0.5cm} Sealed-bid reverse auction-based pricing \cite{lee_2010_sell} & Multiple sellers; a buyer; and an auctioneer & The auctioneer conducts a reverse auction in which sellers submit their asking prices. The auctioneer selects the seller with the lowest asking price as the winner&\begin{itemize}\item {Widely applied to theoretical researches}  \item {Economics: buyer's revenue maximization}  \item{System: data aggregation; resource allocation; and routing protocol} \end{itemize} &Nash equilibrium\tabularnewline \cline{2-5}  
\hline
\vspace{0.5cm} Sealed-bid auction-based pricing \cite{Mihailescu2010Dynamic}& Multiple buyers; a seller; and an auctioneer & The auctioneer conducts an auction in which buyers submit their bidding prices. The auctioneer selects the buyer with the highest bidding price as the winner&\begin{itemize}\item {Widely applied to theoretical researches} \item {Economics: seller's revenue maximization}  \item{System: data aggregation; resource allocation; and routing protocol} \end{itemize} &Competitive equilibrium\tabularnewline \cline{2-5}
\hline
Cost-based pricing \cite{adeel_2014_self}&A seller (or multiple sellers); and a buyer (or multiple buyers) & The seller calculates selling price of an item based on costs charged for the item &\begin{itemize}\item {Widely applied to real world market} \item {Economics: seller's profit maximization}  \item{System: data aggregation; and packet forwarding} \end{itemize}&Optimal solution\tabularnewline \cline{2-5}
\hline
Stackelberg game-based pricing \cite{mei2013}& Multiple sellers, i.e., leaders; multiple buyers, i.e., followers; and multiple brokers & The sellers and brokers determine their own selling prices based on profit functions & \begin{itemize} \item {Applied to real world market and security domain} \item {Economics: sellers' and brokers' profit maximization}  \item{System: relay selection}  \end{itemize} & Stackelberg equilibrium\tabularnewline \cline{2-5}
\hline
\end{tabular}
\par\end{centering}
\end{table*}

\subsection{Auction-Based Pricing for Data Sensing Participation}
\label{subsec:Incentive-based}

Crowdsensing or participatory sensing is a concept to use independent mobile devices to collect and deliver sensing data with the aim to reduce the cost of deploying and maintaining sensor devices. For example, Gigwalk (http://www.gigwalk.com/) provides a marketplace for sensing tasks performed through smartphones. Mobile users, who have installed the Gigwalk client application, can submit their sensing data and receive rewards. However, an important issue is how to set the price that is profitable for the provider and attractive for mobile users. In~\cite{lee_2010_sell}, the authors introduced a reverse auction based dynamic price (RADP) incentive mechanism. The objective is to minimize and stabilize the incentive cost, while achieving sufficient number of participants. In this model, mobile users can sell their sensing data to the provider through their claimed bidding prices. Then, the provider will select users who have the lowest bidding prices. 
The authors demonstrate by simulation that the incentive cost can be reduced by more than 60\% compared with random selection with a fixed price.

\subsection{Cost-Based Pricing for Packet Forwarding}
\label{subsec:Cost_theory}

The authors in~\cite{adeel_2014_self} proposed a cost-based pricing model which can minimize the cost for transmitting packets to the provider through using short-range communications, e.g., Bluetooth and WiFi, with their neighbors instead of transmitting data directly to the provider. After receiving the data, the neighbors can then use the sensing data for themselves or resell it to the provider to gain revenue. To determine the selling price, each device builds a one-hop neighbor table. The device then defines the total cost of sending packets to its neighbors according to remaining energy, resource usage, and costs. The selling prices and corresponding profits are calculated as a function of the total cost, and the device will select a neighbor that has the minimal total cost. Bitmesh (https://www.bitmesh.network/) is one of such data forwarding services that can adopt the aforementioned pricing model. Bitmess allows users to share their Internet connection with peers in a local ``marketplace''. 

\subsection{Pricing Models for Cloud Computing}
\label{subsec:Pricing models}

Due to the flexibility and efficiency, cloud computing becomes a typical infrastructure to store and process a large amount of IoT data collected from devices and sensors. In cloud computing, computing and storage resources can be used in an on-demand basis, reducing total cost of operation. A variety of economic models have been developed for cloud computing services to achieve the highest profits and to meet the user requirements~\cite{Samimi2011Review}. For example, multiple cloud providers can cooperate and share resources to increase scalability and reliability through the federated cloud model. The strategic-proof dynamic pricing (SPDP) scheme~\cite{Mihailescu2010Dynamic} was proposed to improve resource utilization. The SPDP scheme is based on an auction mechanism that the payment for resource allocation is a function of the demand of users and the supply of cloud providers. Such pricing scheme can be applied to practical cloud services such as Zuora (https://www.zuora.com/) that offer cloud resource sharing services.

\subsection{Pricing Models for Information Market}
\label{subsec:pricing_market}

After data is processed, the useful information will be extracted for services to IoT users. Information can be treated as goods that can be sold and bought in a market. In~\cite{mei2013}, the authors studied the information service pricing by formulating a hierarchical game, i.e., a Stackelberg game, among information providers, brokers, and customers. The brokers acquire information from the providers, i.e., leaders, and sell the services to the customers, i.e., followers. The authors also adopted the bundling strategy for selling the information services. The relationship between the information collection cost and willingness-to-pay value of the customers. 

Although there are some SDP schemes applied to IoT systems, few works considered bundled services of multiple providers. Moreover, the service quality due to variable sensors selling data to providers was ignored. Thus, in the next section, we introduce a new pricing scheme for sensing data buying and IoT service subscription with bundling.

\section{Sensing Data Buying and Service Subscription with Bundling}
\label{sec:pricing}

In this section, we introduce a smart data pricing scheme for IoT services that incorporate sensing data buying and service subscription with bundling~\cite{adams18976}. We first discuss the motivations and scenarios. Then, we present the detail of the pricing scheme. The numerical results are presented afterwards. Some works considered the bundling strategy in the smart data pricing, e.g.,~\cite{jin2014}~\cite{wu2014}. Additionally, multi-tier market models were proposed~\cite{li2014} and \cite{zhang2011}. However, the consideration of the bundling strategy in the multi-tier market model taking the unique requirements of IoT services into account was not done before.

\subsection{IoT Services}

\begin{figure}
\begin{center}
$\begin{array}{c} \epsfxsize=4.2 in \epsffile{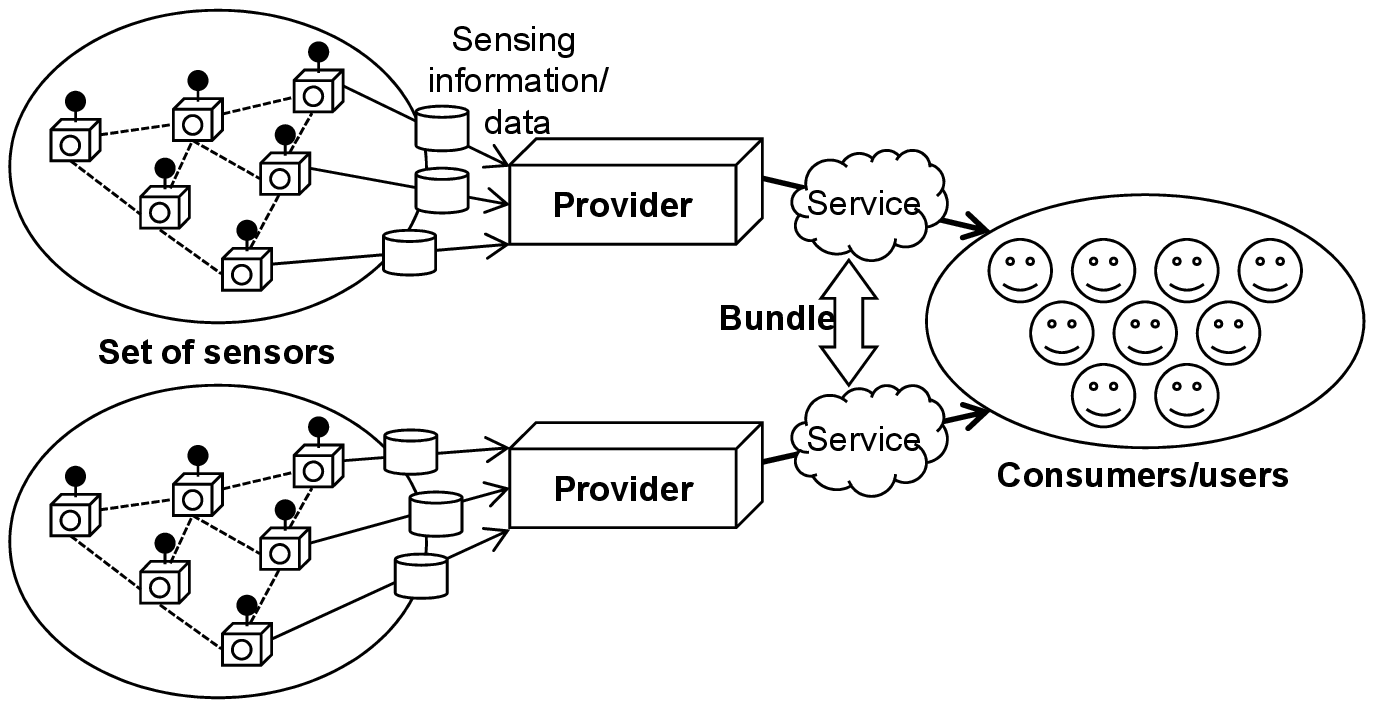} \\ [-0.2cm]
\end{array}$
\caption{Service model of sensing data buying and service subscription with bundling.} 
\label{fig:servicemodel}
\end{center}
\end{figure}

We consider IoT service providers who act as brokers. Each provider buys sensing data from a set of sensors belonged to other owners. The provider then transfers the sensing data and processes it with the purpose to offer a value-added IoT service to a set of consumers or users. The data processing and service delivery can be performed in the facility, i.e., private cloud, of the provider or in the public cloud that the provider rents from other cloud services. Figure~\ref{fig:servicemodel} illustrates the IoT service offered by the provider. There can be multiple providers collecting sensing data from different sets of sensors, and the providers offer the same or different services to the same set of users, i.e., unimodal and multimodal sensing, respectively. We assume that the users need the services from different providers separately with different preferences and values. For example, the sensing data of each provider is from a different geographical area. Thus, the providers do not compete each other. 

This IoT service is inspired and is applicable to many applications and businesses. The following are some pertinent examples.
\begin{itemize}
	\item {\em Placemeter:} Placemeter (https://www.placemeter.com/) is a startup that allows a user to set up a video camera to capture a view from streets and cities and to stream the video data to the Placemeter. The Placemeter then processes it with video analytics. The users will be paid by the Placemeter depending on the view and quality of the video. The video analytics can be used to extract meaningful and useful information to trace crowds and road traffic. Many businesses and government agencies can benefit from the information, e.g., for retailers to open shops and for city agencies to plan public area usage and expansion. 
	\item {\em Road transportation services:} There are a few services that allow drivers to share and obtain road traffic information. Some examples are Google Maps for mobile (http://www.google.com/maps/about/) and Waze (https://www.waze.com/). Drivers simply allow mobile apps on their mobile phones to send related information, e.g., moving speed, to the service provider. Then, the provider translates and processes the information, and informs other drivers of the current road conditions in different areas. 
	\item {\em IoT search engine:} While aforementioned examples are the services for specific purposes, recently IoT search engine services have been introduced. The services provide a generic capability for users to acquire sensing data with various types. Sensor owners can share, exchange, or sell their sensing data to other users. One example is Thingful (https://thingful.net/) that lets sensors be connected and users can browse and obtain desired sensing data. The IoT search engine is able to locate, index, and make sensing data searchable.
\end{itemize}

To develop and operate the IoT services economically and profitably, in the following, we introduce an effective pricing scheme.

\subsection{Pricing Scheme}

As shown in Figure~\ref{fig:servicemodel}, there are three major entities in the IoT service under consideration. 
\begin{itemize}
	\item {\em Sensors} are suppliers of sensing data. Sensor owners sell sensing data to one of the providers. Each sensor has a {\em reservation wage} which is a value of sensing data. In particular, the sensor (owner) will sell its sensing data to a provider if the provider offers to buy with a {\em buying price} greater than the reservation wage. Otherwise, the sensor will not sell the data. The reservation wage can be based on the cost that incurs because of collecting and transmitting the data to the provider.
	\item {\em Users} are the consumers of IoT services offered by the providers. The users can buy any IoT services. Each user has a {\em reservation price} which is a willingness-to-pay value of the IoT service. Specifically, the user will subscribe, i.e., buy, the IoT service from the provider if the provider sells the service with a {\em subscription fee} lower than the reservation price. Otherwise, the user will not subscribe the service. Each user has different reservation prices for different IoT services. The user can also subscribe to multiple IoT services simultaneously.
	\item {\em Providers} buy sensing data from a particular set of sensors, perform value-added data processing, and deliver an IoT service to users. To buy sensing data from the sensors, the provider sets a buying price. Likewise, to sell the IoT service to the users, the provider sets a subscription fee. In addition to setting the price and fee, the providers have an option to cooperate with each other to offer IoT services as a bundle. The objective of the providers is to maximize their profits.
\end{itemize}
Note that the reservation wage and reservation price of the sensors and users, respectively, can be determined from cost-benefit analysis and performance requirements. Alternatively, the conjoint analysis that employs various marketing techniques to quantify such values can also be employed. Moreover, a sensor and user can be physically co-located. For example, vehicle drivers can be data suppliers that transmit their location, traveling speed, and other information to the Google Map or Waze. Likewise, the same drivers, as consumers, can access the Google Map or Waze services to obtain global road network conditions to determine their best driving routes. 

The IoT service providers will maximize their profits through the pricing scheme. Firstly, they have to determine the buying price offered to the sensors and the subscription fee proposed to the users. Secondly, the providers decide whether the providers cooperate to offer their IoT services as a bundle, the bundled subscription fee needs to be optimized. 

\subsubsection{Buying Price and Subscription Fee}
	
We first consider the case without bundling. The provider determines the buying price and subscription fee considering the reservation wage and reservation price of the sensor and user, respectively. 
\begin{itemize}
	\item Firstly, the utility of sensor $i$ is determined by $U_{\mathrm{sen}}(i) = p^{(k)}_{\mathrm{buy}}  - \phi_i$, where $p^{(k)}_{\mathrm{buy}}$ represents the buying price offered by provider $k$, and $\phi_i$ is the reservation wage. The sensor will sell its sensing data if the utility is positive. Given the buying price $p^{(k)}_{\mathrm{buy}}$, the number of sensors $s ( p^{(k)}_{\mathrm{buy}} )$ that sell its data can be determined from $s ( p^{(k)}_{\mathrm{buy}} )  =  \sum_{ i=1}^I {\mathbf{1}}_{ U_{\mathrm{sen}}(i) > 0} $, where ${\mathbf{1}}_{ U_{\mathrm{sen}} (i) > 0 } $ is an indicator function that returns one if the utility of sensor $i$ is greater than zero, and zero otherwise. Here, $I$ is the number of potential sensors.
	\item Secondly, the utility of user $j$ subscribing to provider $k$ is expressed as $U_{\mathrm{usr}}(j,k) = Q_k(s) \theta_{j,k} - p^{(k)}_{\mathrm{fee}}$, where $Q_k(s)$ is the service quality, which is defined as an increasing function of the number of sensors $s$ participating in the IoT service of provider $k$, $\theta_{j,k}$ is the maximum reservation price, and $p^{(k)}_{\mathrm{fee}}$ is the subscription fee. We assume that the service quality varies between 0 and 1. Naturally, when there are more sensing data from more sensors, the quality of IoT service tends to be better. Thus, the users will appreciate more from the service. For example, if more drivers supply their driving status, it is likely that the estimated road traffic information will be more accurate. Here, $Q_k(s) \theta_{j,k}$ is basically a reservation price of the user. Similar to the sensor, the user will subscribe the service if the utility is positive.
\end{itemize}
The profit of provider $k$ is 
\begin{equation}
	F_k = \underset{\mathrm{Revenue}} {\underbrace{  p^{(k)}_{\mathrm{fee}} \sum_{ j = 1 }^J  {\mathbf{1}}_{U_{\mathrm{usr}}(j,k) > 0} } }
			 - \underset{\mathrm{Cost}} {\underbrace{	p^{(k)}_{\mathrm{buy}} s ( p^{(k)}_{\mathrm{buy}} )	} } 	,
\end{equation}
where ${\mathbf{1}}_{U_{\mathrm{usr}}(j,k) > 0}$ is an indicator function that returns one if the utility of user $j$ is greater than zero, and zero otherwise. $J$ is the total number of potential users. Thus, the optimal buying price and subscription fee are obtained from $( p^{(k)*}_{\mathrm{fee}}, p^{(k)*}_{\mathrm{buy}} ) = \arg \max_{ ( p^{(k)}_{\mathrm{fee}}, p^{(k)}_{\mathrm{buy}} ) } F_k ( p^{(k)}_{\mathrm{fee}}, p^{(k)}_{\mathrm{buy}}  )$.

\subsubsection{Bundled Service}

A set or a coalition of providers, denoted by ${\mathcal{K}}$, can cooperate to offer their IoT services as a bundle with the fee $p_{\mathrm{bun}}$. They can self-organize the coalitions or the third party can facilitate the formation. For example, an IoT search engine can let subscribers, who pay the fee, access all sensing data such as the weather and road traffic conditions. The cooperative providers will jointly optimize the buying prices and subscription fee of the bundle. While the utility of a sensor is the same as that in the case without bundling, the utility of a user to a bundle is defined as $U_{\mathrm{usrb}}(j) = \sum_{k \in {\mathcal{K}} } Q_k(s) \theta_{j,k} - p_{\mathrm{bun}}$. Basically, the user will buy the bundle if the total reservation price of all services in the bundle is higher than the bundle subscription fee.

The profit of the coalition of providers is 
\begin{equation}
	F_{\mathcal{K}} =   p_{\mathrm{bun}} \sum_{ j = 1 }^J  {\mathbf{1}}_{U_{\mathrm{usrb}}(j) > 0}  
			 - \sum_{k \in {\mathcal{K}} } p^{(k)}_{\mathrm{buy}} s ( p^{(k)}_{\mathrm{buy}} ) .
\end{equation}
Similarly, the buying prices and bundle subscription fee are optimized to maximize the profit. Notably, $F_{\mathcal{K}}$ is the profit of every provider in coalition ${\mathcal{K}}$. Thus, a fair profit sharing scheme is needed. The scheme must ensure that all providers gain their profit higher than that without joining a coalition, i.e., making a bundle. From the cooperative game theory literature, the solution concepts such as a Shapley value and Nash bargaining solution~\cite{cooperativegamebook} can be applied. 

\subsection{Numerical Examples}

To simplify the presentation of the numerical analysis, we use the following simple setting. There are two sets of sensors, and thus two providers offering services 1 and 2. The reservation wage of the sensors and the reservation price of users are uniformly distributed between 0 and 1. The number of sensors in each set is 50, and the number of users is 200. The service quality function of a set of sensors is logarithmic, i.e., $Q(s) = q \log ( 1 + s/I )$, where $s$ is the average number of participated sensors. $I$ is the total number of sensors in a set. $q$ is a sensing quality factor to be varied. A concave quality function is reasonable in the sense that the improvement of the service quality becomes diminishing when there are more sensors.

\begin{figure}
\begin{center}
$\begin{array}{c} \epsfxsize=3.6 in \epsffile{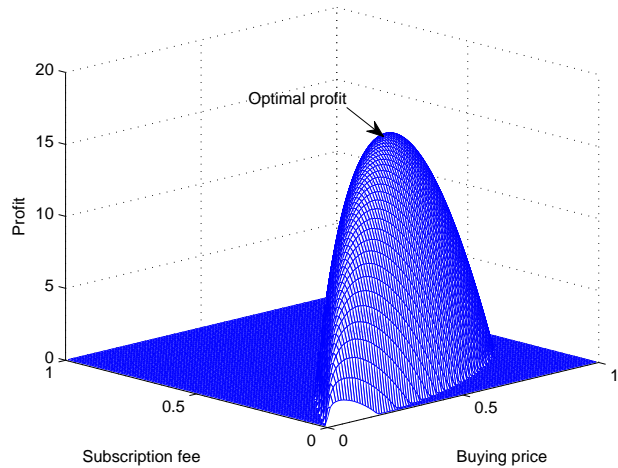} \\ [-0.2cm]
\end{array}$
\caption{Profit of a service without bundling.} 
\label{fig:result_profit}
\end{center}
\end{figure}

We first show the impact of the sensing data buying price and service subscription fee to a profit of one service provider. The provider buys sensing data from one set of sensors and sells a service from processing the sensing data to the users. Figure~\ref{fig:result_profit} shows the profit of the provider. Apparently, there are an optimal sensing data buying price and service subscription fee that maximize the profit. The profit is a unimodal function. Thus, numerical methods, e.g., a simplex method, can be applied to obtain the optimal solutions. In this case, when the buying price is low, the service quality is low due to few sensors selling data, i.e., low supply. Thus, the provider cannot charge a high subscription fee to the users and cannot gain a high profit. If the buying price is too high, the service quality improves, but the cost increases, and the profit plunges. Similarly, when the subscription fee is low, the revenue of the provider is small. However, when the subscription fee is too high, few users will purchase the service, i.e., low demand. Consequently, the revenue and profit tumble. 

\begin{figure}
\begin{center}
$\begin{array}{cc} 
\epsfxsize=3.6 in \epsffile{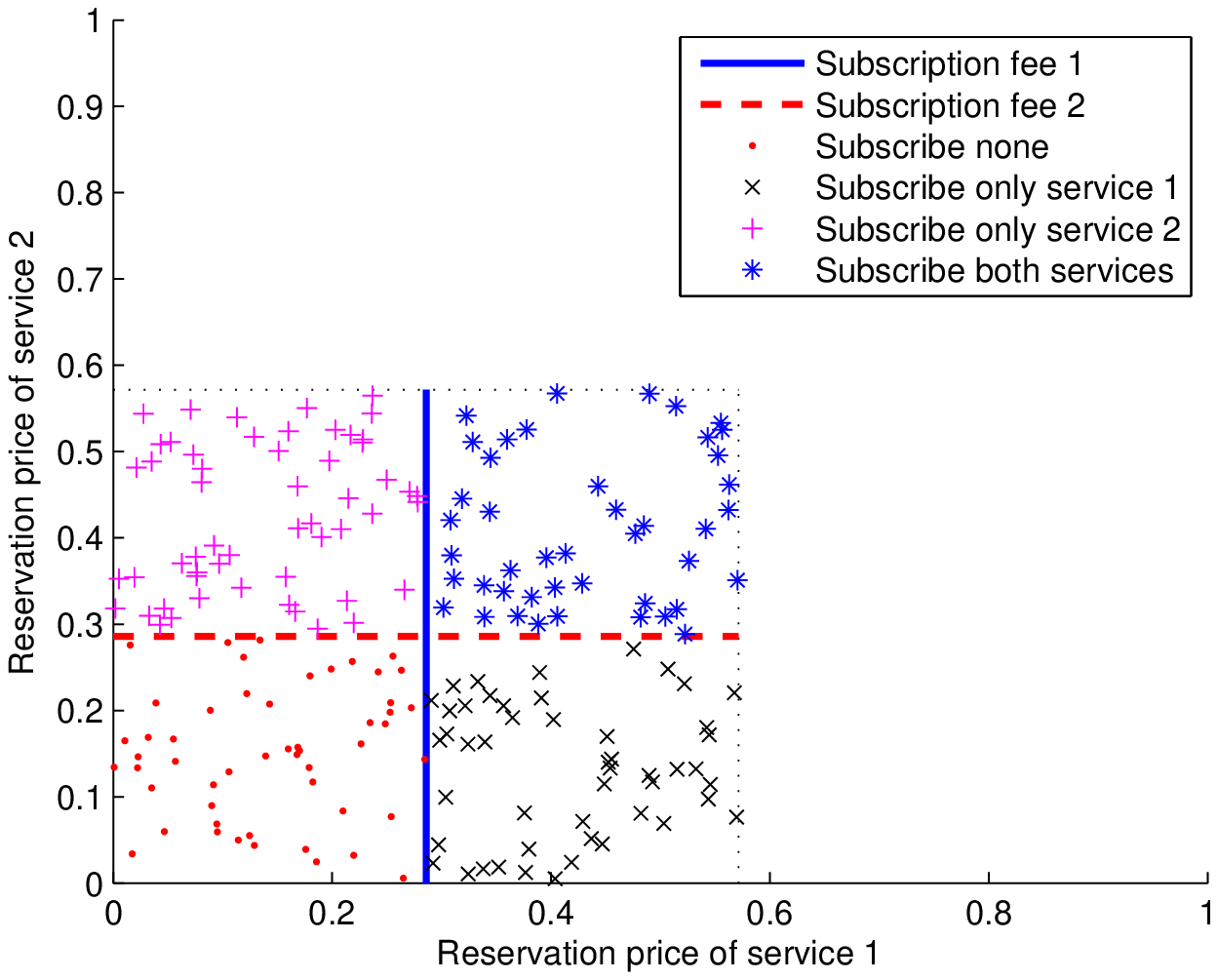} &
\epsfxsize=3.6 in \epsffile{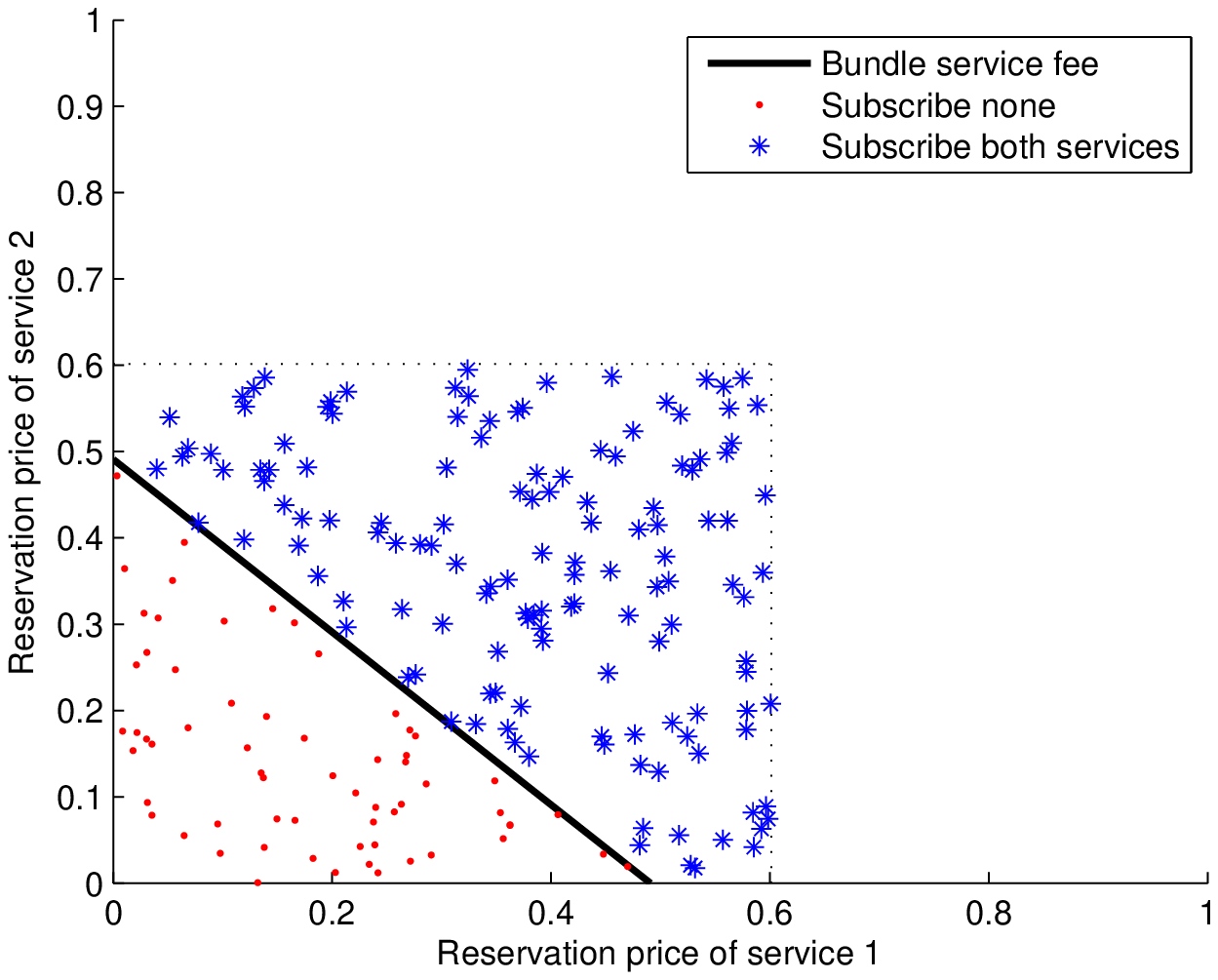} \\ 
(a)	&	(b)		\\	[-0.2cm]
\end{array}$
\caption{Subscription fees of (a) selling services separately and (b) selling services as a bundle under different customer reservation prices.}
\label{fig:result_bundle}
\end{center}
\end{figure}

We then consider the impact of the bundle. Here, we consider a symmetric setting of both providers for simplicity. Figures~\ref{fig:result_bundle}(a) and (b) show the optimal subscription fees when two providers offer their services separately and as a bundle, respectively. The locations of the markers in the figures correspond to the reservation prices of users. For selling service separately (Figure~\ref{fig:result_bundle}(a)), the users will subscribe a service if their reservation price is higher than each of subscription fee. Here, the optimal sensing data buying price from both sets of sensors is 0.486 which leads to the maximum reservation price of 0.571. The optimal service subscription fees for both providers are 0.286. Since the services are sold separately, the optimal fees are shown as vertical and horizontal straight lines for providers 1 and 2, respectively. Thus, there are four regions. Users subscribe no service if none of their reservation price is higher than the fees. Users subscribe one of the services if one of their reservation prices is greater than the fee. Finally, users subscribe both the services if all their reservation prices are higher than the fees.

For selling service as a bundle (Figure~\ref{fig:result_bundle}(b)), the users subscribe both services if the sum of their reservation prices is higher than the subscription fee of the bundled services. Here, the optimal sensing data buying price from both sets of sensors is 0.517 which results in the maximum reservation prices of 0.601. The optimal bundled service subscription fee is 0.491. Thus, there are two regions that correspond to the users subscribing and not subscribing the bundled services.

From Figures~\ref{fig:result_bundle}(a) and (b), the profit of selling the bundled services is higher than that of selling services separately, i.e., 38.723 versus 33.524. This can be observed from that more users subscribe both services in Figure~\ref{fig:result_bundle}(b) than that in Figure~\ref{fig:result_bundle}(a). Additionally, selling services as a bundle allows providers to charge a higher subscription fee.

\begin{figure}
\begin{center}
$\begin{array}{c} \epsfxsize=3.6 in \epsffile{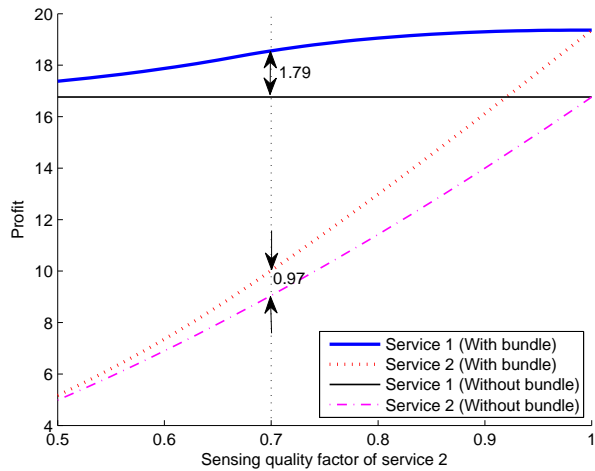} \\ [-0.2cm]
\end{array}$
\caption{Profit of services under different sensing quality of service 2.}
\label{fig:result_sq}
\end{center}
\end{figure}

Next, we consider an asymmetric case that the sensing quality factor for the set of sensors for service 2 is varied, while that of provider 1 is fixed at one. We apply the Shapley value for profit sharing between two service providers. Figure~\ref{fig:result_sq} shows the profits obtained by the two providers when they cooperate and do not cooperate. Clearly, when they cooperate to sell their services as a bundle, the individual profits are higher than that without cooperation, i.e., selling services separately. We can observe that the gains from the cooperation for both providers are different. This implies that the providers can tolerate the integration cost differently. Here, the integration cost incurs because of the cooperation and bundling. Consider the sensing quality factor at 0.7 for service 2. The gain for provider 2 is smaller than that of provider 1, i.e., 0.97 versus 1.79. Thus, if the integration cost to provider 2 is more than 0.97, then provider 2 will not want to sell its service as a bundle. Likewise, if the integration cost to provider 1 is more than 1.79, then provider 1 will not cooperate. 

Based on the proposed sensing data buying and service subscription with bundling, the following points can be considered for the future work.
\begin{itemize}
	\item {\em Strategic sensors and users:} Sensors and users can adjust their reservation wage and reservation price, respectively, based on market conditions. In this case, an auction can be one of the suitable tools to determine the equilibrium reservation wage and reservation price. The discriminatory pricing scheme can be developed. 
	\item {\em Cooperation and collusion:} To buy sensing data from sensors and to sell services to users, providers can compete or cooperate with each other. In a competitive environment, the provider will set the buying price and the subscription fee to maximize its individual profit given the strategies of other providers. A Nash equilibrium solution can be adopted. Alternatively, the providers can collude to maximize their profits collectively. The collusion formation and prevention in the market can be studied.
	\item {\em Quality of data:} The providers can adjust the buying price for each sensor individually to encourage it to supply high quality sensing data. In particular, the sensors can optimize data quality based on the buying price and their resource usage for collecting, processing, and transmitting data to the provider. For example, a camera as a sensor can supply higher video quality, but at the cost of more energy consumption and bandwidth usage. Hence, the camera will do so only when the buying price is sufficiently high. Joint pricing and performance optimization models can be developed.
\end{itemize}


\section{Conclusion}
\label{sec:conclusion}

In this article, we have considered smart data pricing (SDP) for IoT systems and services. We have first introduced an overview of IoT including its architecture, benefits, and business models. Then, we have reviewed some related work of applying SDP to IoT. We have proposed a new pricing scheme for IoT service providers taking into account sensing data buying and subscription with bundling. The numerical results have clearly shown that with bundling, multiple providers can form a coalition to achieve higher profit.


\begin{thebibliography}{99}


\bibitem{Gubbi2013Internet}
J.~Gubbi, R.~Buyya, S.~Marusic, and M.~Palaniswami, ``Internet of Things (IoT): A vision, architectural elements, and future directions,'' \emph{Future Generation Computer Systems}, vol. 29, no. 7, pp. 1645-1660, September 2013.

\bibitem{Bucherer2011Business}
E.~Bucherer and D.~Uckelmann, ``Business models for the Internet of things,'' in \emph{Architecting the internet of things}, pp. 253-277, Springer Berlin Heidelberg, 2011. 

\bibitem{Sen_2014_Smart}
S.~Sen, C.~J.~Wong, S.~Ha, M.~Chiang, ``Smart data pricing: Economic solutions to network congestion,'' \emph{Communications of the ACM}, 2014.

\bibitem{IEEE_IoT_definition}
R.~Minerva, A.~Biru, and D.~Rotondi, ``Towards a definition of the Internet of Things (IoT),'' \emph{IEEE Technical Report}, Revision 1, Published 27 May 2015. 

\bibitem{Stankovic2014}
J. A. Stankovic, ``Research Directions for the Internet of Things,'' {\em IEEE Internet of Things Journal}, vol. 1, no. 1, pp. 3-9, February 2014.

\bibitem{Osterwalder2010Business}
A.~Osterwalder, and Y.~Pigneur, ``Business model generation: a handbook for visionaries, game changers, and challengers,'' \emph{John Wiley \& Sons}, 2010.


\bibitem{lee_2010_sell}
J.~S. Lee and B.~Hoh, ``Sell your experiences: a market mechanism based incentive for participatory sensing,'' in \emph{IEEE International Conference on Pervasive Computing and Communications}, pp.~60-68, 2010.

\bibitem{Samimi2011Review}
P.~Samimi and A.~Patel, ``Review of pricing models for grid \& cloud computing,'' in \emph{IEEE Symposium on Computers \& Informatics}, pp. 634-639, Kuala Lumpur, Malaysia, March 2011. 

\bibitem{Mihailescu2010Dynamic}
M.~Mihailescu and Y.~M.~Teo, ``Dynamic resource pricing on federated clouds,'' in \emph{IEEE/ACM International Conference on Cluster, Cloud and Grid Computing}, pp. 513-517, Melbourne, Australia, May 2010.

\bibitem{adeel_2014_self}
U.~Adeel, S.~Yang, and J.~A. McCann, ``Self-optimizing citizen-centric mobile urban sensing systems,'' in \emph{Proceedings of the 11th International Conference on Autonomic Computing}, pp. 161-167, 2014.

\bibitem{mei2013}
L. Mei, W. Li, and K. Nie, ``Pricing decision analysis for information services of the Internet of things based on Stackelberg game,'' in {\em Proceedings of International Conference on Logistics, Informatics and Service Science (LISS)}, Springer, pp. 1097-1104, 2013.


\bibitem{adams18976}
W. J. Adams and J. L. Yellen, ``Commodity bundling and the burden of monopoly,'' {\em The Quarterly Journal of Economics}, vol. 90, no. 3, pp. 475-498, August 1976.

\bibitem{bakos1999}
Y. Bakos and E. Brynjolfsson, ``Bundling information goods: Pricing, profits, and efficiency,'' {\em Management Science}, vol. 45, no. 12, pp. 1613 - 1630, December 1999.


\bibitem{jin2014}
Y. Jin and Z. Pang, ``Smart data pricing: To share or not to share?,'' in {\em Proceedings of IEEE Conference on Computer Communications Workshops (INFOCOM WKSHPS)}, pp. 583-588, April-May 2014.

\bibitem{wu2014}
W. Wu, R. T. B. Ma, and J. C. S. Lui, ``Exploring bundling sale strategy in online service markets with network effects,'' in {\em Proceedings of IEEE INFOCOM}, pp. 442-450, April-May 2014.

\bibitem{le2014}
T. Le, M. Beluri, M. Freda, J.-L. Gauvreau, S. Laughlin, and P. Ojanen, ``On a new incentive and market based framework for multi-tier shared spectrum access systems,'' in {\em Proceedings of IEEE International Symposium on Dynamic Spectrum Access Networks (DYSPAN)}, pp. 477-488, 1-4 April 2014.

\bibitem{zhang2011}
X. Zhang and B. Li, ``On the Market Power of Network Coding in P2P Content Distribution Systems,'' {\em IEEE Transactions on Parallel and Distributed Systems}, vol. 22, no. 12, pp. 2063-2070, December 2011.


\bibitem{cooperativegamebook}
B. Peleg and P. Sudholter, {\em Introduction to the Theory of Cooperative Games}, Springer, October 2007.



\end{thebibliography}
\end{document}